\documentstyle[12pt,titlepage,epsfig]{article}

\font\grande=cmr9.5 scaled \magstep4
\font\medio=cmr9.5 scaled \magstep2
\outer\def\beginsection#1\par{\medbreak\bigskip
      \message{#1}\leftline{\bf#1}\nobreak\medskip
\vskip-\parskip
      \noindent}

\def\laq{\raise 0.4ex\hbox{$<$}\kern -0.8em\lower 0.62
ex\hbox{$\sim$}}
\def\gaq{\raise 0.4ex\hbox{$>$}\kern -0.7em\lower 0.62
ex\hbox{$\sim$}}

\begin{document}
\bibliographystyle {unsrt}

\titlepage

\begin{flushright}
CERN-PH-TH/2005-068
\end{flushright}

\vspace{15mm}
\begin{center}
{\grande Interacting viscous mixtures}\\
\vspace{15mm}
 Massimo Giovannini\footnote{e-mail address: massimo.giovannini@cern.ch}\\
\vspace{8mm}
{\sl Centro ``Enrico Fermi", Compendio del Viminale, Via 
Panisperna 89/A, 00184 Rome, Italy}\\
\vspace{8mm}
{\sl Department of Physics, Theory Division, CERN, 1211 Geneva 23, Switzerland}

\end{center}

\vskip 2cm
\centerline{\medio  Abstract}

\noindent
Gravitational and hydrodynamical perturbations 
are analysed in a relativistic plasma 
containing a mixture of interacting fluids characterized by a 
non-negligible bulk viscosity coefficient.  The energy-momentum 
transfer between the cosmological fluids, as well as the fluctuations of the 
bulk viscosity coefficients, are analyzed simultaneously
with the aim of deriving a generalized set of evolution equations 
for  the entropy and curvature fluctuations. For 
typical length scales larger than the Hubble radius, 
the fluctuations of the bulk viscosity coefficients 
and of the decay rate provide source terms for the 
evolution of both the curvature and the entropy fluctuations. 
According to the functional dependence of the bulk 
viscosity coefficient on the energy densities 
of the fluids composing the system, the mixing of entropy and 
curvature perturbations is scrutinized both analytically and numerically.
\vspace{5mm}
\vfill
\newpage
If a relativistic plasma contains a mixture of inviscid fluids with 
negligible transfer of energy and momentum, the evolution of
entropy fluctuations is characterized, in the long-wavelength limit, 
by the absence of source terms containing curvature perturbations. 
This property has relevant consequences on the dynamics 
of the coupled system of gravitational and hydrodynamical 
perturbations. It implies, for instance, that curvature 
perturbations are conserved, in the long-wavelength limit,
under rather general assumptions \cite{KS,bst,weinberg}. Long-wavelength 
fluctuations in the spatial curvature determine, via the Sachs--Wolfe 
effect, the temperature inhomogeneities observed in the 
microwave sky (see, for instance, \cite{maxrev}). 

Mixtures of relativistic fluids are a useful toy model 
that can be investigated with the purpose of inferring some general 
properties of the evolution equations of curvature and entropy 
fluctuations. Moreover, multifluid systems are per se relevant
to the model-independent discussion of initial conditions of CMB 
anisotropy \cite{KS,maxrev}: for instance, the five isocurvature 
modes supported by the predecoupling plasma may be discussed, in their 
simplest realization, by a truncated Einstein--Boltzmann system of equations 
whose lower multipole moments reproduce indeed a multifluid 
hydrodynamical description \cite{turok}.

One of the assumptions often invoked in the analysis of multifluid 
systems is that the bulk viscosity coefficient and its possible 
spatial variation have a negligible impact on the dynamics. While 
this assumption may be justified in some specific system, it may 
not be true in the early stages of the life of the Universe (see, for instance, \cite{barrow1,barrow2,barrow3}).  
Unlike other dissipative effects, the presence of  bulk viscosity 
does not spoil the isotropy of the background geometry. 
Therefore, consider a mixture of two relativistic fluids 
(the a-fluid and the b-fluid) obeying a set of generally covariant 
evolution equations  formed by the Einstein equations \footnote{Units of
$8\pi G = 1$ will be used throughout. Notice, to avoid confusions, that the Latin (lower-case roman) subscripts a, b, c d, ... will denote, in the present paper, different fluids present in the relativistic plasma.
Greek (lower-case) subscripts will denote tensor indices. Latin (lower-case italic) subscripts $i,j,k, ...$ will denote the spatial components of a tensor. }
\begin{equation}
R_{\mu}^{\nu} - \frac{1}{2} \delta_{\mu}^{\nu} R = \frac{1}{2} {\cal T}_{\mu}^{\nu}
\label{E1}
\end{equation}
and by the evolution equations of the energy-momentum 
tensors of each fluid of the mixture, i.e. 
\begin{eqnarray}
&& \nabla_{\mu} {\cal T}^{\mu\nu}_{\rm a} = - \Gamma g^{\nu\beta} 
u_{\beta}(p_{\rm a} + \rho_{\rm a}),
\label{eqTmna}\\
&&   \nabla_{\mu} {\cal T}^{\mu\nu}_{\rm b} = \Gamma g^{\nu\beta} 
u_{\beta}(p_{\rm a} + \rho_{\rm a}),
\label{eqTmnb}
\end{eqnarray}
where $u_{\beta}$ is the total velocity field of the mixture.
Equations (\ref{eqTmna}) and (\ref{eqTmnb}) describe the situation 
where the a-fluid decays into the b-fluid with decay rate $\Gamma$. It is 
evident from the form of Eqs. (\ref{eqTmna}) and (\ref{eqTmnb}) 
that the total energy-momentum tensor of the mixture, i.e. 
${\cal T}^{\mu\nu} = {\cal T}^{\mu\nu}_{\rm a} + {\cal T}^{\mu\nu}_{\rm b}$ 
is covariantly conserved, i.e. $\nabla_{\mu} {\cal T}^{\mu\nu}=0$.
The total energy-momentum tensor of each species 
is given by the sum of an inviscid contribution, denoted by 
$T^{\mu\nu}_{\rm a,\,b}$ and by a viscous contribution, denoted by 
$\tilde{T}^{\mu\nu}_{\rm a,\,b}$, i.e. 
\begin{eqnarray}
&& {\cal T}_{\rm a,\,b}^{\mu\nu} = T_{\rm a,\,b}^{\mu\nu} + 
\tilde{T}_ {\rm a,\,b}^{\mu\nu},
\label{ab1}\\
&& T_{{\rm a,\,b}}^{\mu\nu} = ( p_{\rm a,\,b} + \rho_{\rm a,\,b}) 
u_{\rm a,\,b}^{\mu} u_{\rm a,\,b}^{\nu} - p_{\rm a,\,b} g^{\mu\nu},
\label{ab2}\\
&& \tilde{T}_{{\rm a,\,b}}^{\mu\nu} = \xi_{\rm a,\,b} \biggl( g^{\mu\nu} 
- u_{\rm a,\,b}^{\mu} u_{\rm a,\,b}^{\nu} \biggr) \nabla_{\alpha}u^{\alpha}_{\rm a,\,b},
\label{ab3}
\end{eqnarray}
where the subscript in the various fluid quantities simply means that 
Eqs. (\ref{ab1}), (\ref{ab2}) and (\ref{ab3}) hold, independently, for the 
a- and b-fluids. So, for instance, in
 Eqs. (\ref{ab2}) and (\ref{ab3}), $u_{\rm a}^{\mu}$ and $u_{\rm b}^{\mu}$ 
 denote the peculiar velocities of each fluid of the mixture.

In a spatially flat metric of Friedmann--Robertson--Walker (FRW) type
charaterized by a background line element 
\begin{equation}
d s^2 = \overline{g}_{\mu\nu} d x^{\mu} d x^{\nu} =
a^2(\tau) [ d\tau^2 - d\vec{x}^2],
\end{equation}
Eqs. (\ref{eqTmna}) and (\ref{eqTmnb}) imply 
\begin{eqnarray}
&& \rho_{\rm a}' + 3 {\cal H}
 ( \rho_{a} + {\cal P}_{\rm a}) + a\overline{\Gamma} ( \rho_{\rm a} + p_{\rm a}) =0,
\label{bconsa}\\
&& \rho_{\rm b}' + 3 {\cal H} ( \rho_{b} + {\cal P}_{\rm b}) -
 a\overline{\Gamma} ( \rho_{\rm a} + p_{\rm a}) =0,
 \label{bconsb}
 \end{eqnarray}
 where the prime denotes a derivation with respect to the 
 conformal time coordinate $\tau$ and ${\cal H} =a'/a$.
In Eqs. (\ref{bconsa}) and (\ref{bconsb}), ${\cal P}_{\rm a,\, b}$ 
denote the total effective pressure of each species, i.e. 
\begin{equation}
{\cal P}_{\rm a,\,b} = p_{\rm a,\,b} - 3 \frac{\cal H}{a} 
\overline{\xi}_{\rm a,\,b},
\label{calpdef}
\end{equation}
while $p_{\rm a,\,b}$ denote the inviscid pressures of each species.
In Eq. (\ref{calpdef}), $\overline{\xi}_{\rm a,\,b}$ denote 
the bulk viscosity coefficient evaluated on the background geometry.
As  will be discussed later, the bulk viscosity coefficient may depend 
on both $\rho_{\rm a}$ and $\rho_{\rm b}$.
Equations (\ref{bconsa}) and (\ref{bconsb}) lead to the evolution 
 of the total energy and pressure  densities 
\begin{equation}
\rho' + 3 {\cal H} ( \rho + {\cal P}) =0,
\label{CB1}
\end{equation}
where $\rho = \rho_{\rm a} + \rho_{\rm b}$ and 
${\cal P} = {\cal P}_{\rm a} + {\cal P}_{\rm b}$.
Equations (\ref{bconsa}) and (\ref{bconsb}) must be supplemented 
by the explicit background form of Eq. (\ref{E1}), i.e. 
\begin{eqnarray}
&&  3 {\cal H}^2 =  a^2 \rho,
\label{Eb1}\\
&& 2({\cal H}^2 - {\cal H}') =  a^2 ( \rho + {\cal P}),
\label{Eb2}
\end{eqnarray}
where, again, $(\rho + {\cal P})$ is the total effective enthalpy 
that contains the background viscosity coefficient 
of the  mixture $\overline{\xi} = \overline{\xi}_{\rm a} + \overline{\xi}_{\rm b} $.

We are now interested in deriving the evolution of the entropy 
and total-curvature fluctuations of the system. 
Both the entropy perturbations and 
the perturbations in the total spatial curvature can be written in terms 
of $\zeta_{\rm a}$ and $\zeta_{\rm b}$ which are related in the 
 off-diagonal gauge \cite{maxoff} (see also \cite{hoff,hwang}) to the 
density contrasts of the individual fluids of the mixture \cite{malikwands}:
\begin{eqnarray}
&& {\cal S} = - 3 (\zeta_{\rm a} - \zeta_{\rm b}),
\label{entrdef}\\
&& \zeta = \frac{\rho_{\rm a}'}{\rho'} \zeta_{\rm a} + 
 \frac{\rho_{\rm b}'}{\rho'} \zeta_{\rm b}.
\label{curvdef}
\end{eqnarray}
 In the following we are going 
to exploit the off-diagonal gauge \cite{maxoff} (also called 
uniform-curvature gauge \cite{hoff}),  which is particularly convenient 
for the problem at hand. The results will be 
exactly the same of those obtainable in the framework 
of gauge-independent descriptions (see \cite{maxlon}). In fact, 
the quantities
${\cal S}$ and $\zeta$, defined in terms of $\zeta_{\rm a}$ and $\zeta_{\rm b}$, 
are invariant under infinitesimal coordinate transformations. Consequently they 
can be computed in any suitable (non-singular) coordinate system.

Concerning Eq. (\ref{curvdef}) it could be noticed 
that the gauge-invariant definition of the spatial curvature 
perturbation can be slightly different for wavelengths 
smaller than the Hubble radius. However, since in the problem 
at hand we are mainly interested in super-Hubble fluctuations, 
Eq. (\ref{curvdef}) is numerically equivalent to the 
curvature inhomogeneities defined, more conventionally, 
 from the curvature 
fluctuations on comoving orthogonal hypersurfaces \cite{bst,bp,lyth} (see also 
\cite{maxrev}).
 The physical interpretation of the entropy fluctuations defined 
 in Eq. (\ref{entrdef}) can be understood, for instance, in the case 
 of an inviscid mixture of cold dark matter (CDM) and radiation. 
 In this situation the a-fluid 
 is given by pressureless matter and the b-fluid by radiation. Applying 
 the covariant conservation equations for two (inviscid and non-interacting) species, it is clear that ${\cal S}$ is nothing but 
 the fractional fluctuation in the specific entropy $\varsigma = T^3/n_{\rm CDM}$ (where $T$ is the radiation temperature and $n_{\rm CDM}$ is the 
 CDM number density), i.e. the entropy density per CDM particle:
 \begin{equation}
 {\cal S} \equiv \frac{\delta \varsigma}{\varsigma} = \delta_{\rm CDM} - 
 \frac{3}{4} \delta_{\rm r},
 \end{equation}
 where $\delta_{\rm CDM}$ and $\delta_{\rm r}$ are the density 
 contrasts for the CDM and for the radiation fluids.  
 
 In the off-diagonal gauge the spatial components 
of the perturbed metric vanish and, hence, the only 
components of the perturbed line element are:
\begin{equation}
\delta g_{00} = 2 a^2 \phi,\,\,\,\,\,\,\,\,\, \delta g_{0i} = - a^2 \partial_{i} B.
\end{equation}
Since, in the long-wavelength limit, 
$({\cal H}' - {\cal H}^2) \phi = {\cal H}^2 \zeta$, it turns 
out that in the off-diagonal  gauge, $\delta g_{00}$ is connected to $\zeta$.  
As anticipated, $\zeta_{\rm a}$ and $\zeta_{\rm b}$ can be 
expressed in terms of the fluctuations of the density contrasts of the individual 
fluids, i.e. 
\begin{equation}
\zeta_{\rm a} = - \frac{\cal H \,\rho_{\rm a}}{\rho_{\rm a}'}\delta_{\rm a},
\,\,\,\,\,\,\,\,\,\,\,\,\,\,\,\,\,\,\zeta_{\rm b} = - \frac{\cal H \,\rho_{\rm b}}{\rho_{\rm b}'}\delta_{\rm b}.
\label{ztod}
\end{equation}
The evolution equations obeyed by the density contrasts $\delta_{\rm a} = 
\delta \rho_{\rm a}/\rho_{\rm a}$ and $\delta_{\rm b} = 
\delta\rho_{\rm b} /\rho_{\rm b}$ are derived 
by perturbing Eqs. (\ref{eqTmna}) and (\ref{eqTmnb}) to first order 
in the amplitude of the metric and hydrodynamical fluctuations:
\begin{eqnarray}
&& \delta_{\rm a}' +
 ( 3 {\cal H} + a \overline{\Gamma})(c_{\rm s\,a}^2 - w_{\rm a}) \delta_{\rm a} 
 + \frac{9 {\cal H}^2}{a \rho_{\rm a}} [ \overline{\xi}_{\rm a} ( \phi + \delta_{\rm a} ) - \delta \xi_{\rm a}] +  a (1 + w_{\rm a}) 
 \overline{\Gamma} ( \delta_{\Gamma} + \phi)
 \nonumber\\ 
 &&  +  \biggl[ ( 1 + w_{\rm a}) - 6 \frac{ {\cal H}\,\overline{\xi}_{\rm a}}{a\, \rho_{\rm a}}\biggr] \theta_{\rm a} =0 ,
 \label{dca}\\
 && \delta_{\rm b}' + 3 {\cal H}(c_{\rm s\, b}^2 - w_{\rm b}) \delta_{\rm b}
 + a\overline{\Gamma} \frac{\rho_{\rm a}}{\rho_{\rm b}} [ ( 1 + w_{\rm a}) 
( \delta_{\rm b} -\delta_{\Gamma} -\phi) - ( 1 + c_{\rm s\, a}^2 ) \delta_{\rm a}]
  \nonumber\\
&&  + \frac{ 9 {\cal H}^2 }{ a \rho_{\rm b}} [ \overline{\xi}_{\rm b} ( \phi + 
 \delta_{\rm b}) - \delta \xi_{\rm b}] 
  + \biggl[ ( 1 + w_{\rm b}) - 6 \frac{ {\cal H} \,\overline{\xi}_{\rm b}}{a \rho_{\rm b}} \biggr] \theta_{\rm b} =0.
\label{dcb}
 \end{eqnarray}
Concerning Eqs. (\ref{dca}) and (\ref{dcb}) 
a few comments are in order:
\begin{itemize}
\item{} for notational convenience the barotropic indices (i.e. 
$w_{\rm a}$, $w_{\rm b}$) and the sound speeds 
(i.e. $c_{\rm s\, a}^2$ and $c_{\rm s\,b}^2$) have been introduced 
for the inviscid component of each species of the plasma; 
if the inviscid component is parametrized in terms of a perfect 
relativistic fluid $c_{\rm s\, a,\,b}^2 \equiv w_{\rm a,\, b}$;
\item{} $\delta_{\Gamma} = \delta \Gamma/\overline{\Gamma}$ 
is the fractional fluctuation of the decay rate computed 
in the off-diagonal gauge;
\item{} $\delta\xi_{\rm a}$ and $\delta\xi_{\rm b}$ 
denote the fluctuations of the bulk viscosity coefficients; 
later on it will also be convenient to introduce  
the fluctuation in the total viscosity, i.e. $\delta\xi = \delta \xi_{\rm a} + 
\delta \xi_{\rm b}$;
\item{} finally, $\theta_{\rm a} = \partial_{i} v^{i}_{\rm a} = \nabla^2 v_{\rm a}$
and $\theta_{\rm b} = \partial_{i} v^{i}_{\rm b} = \nabla^2 v_{\rm b}$ are 
the divergences of the peculiar velocity field of each species; note 
that the global velocity $\theta = \partial_{i} v^{i}$ field (with 
$\delta {\cal T}_{0}^{i} = ( \rho + {\cal P}) v^{i}$) is 
recovered from $\theta_{\rm a}$ and $\theta_{\rm b}$ by recalling that $(p+ \rho)\theta = (p_{\rm a} + \rho_{\rm a})\theta_{\rm a} 
+ (p_{\rm b} + \rho_{\rm b}) \theta_{\rm b}$.
\end{itemize}

Equations (\ref{dca}) and (\ref{dcb}) must be supplemented by the perturbed components of Eq. (\ref{E1}); in particular by the Hamiltonian and momentum 
constraints: 
\begin{eqnarray}
&&  {\cal H} \nabla^2 B + 3 {\cal H}^2 \phi + \frac{a^2}{2} \delta \rho =0,
\label{ODham}\\
&& \nabla^2 [ {\cal H} \phi + ({\cal H}^2 - {\cal H}') B] + \frac{a^2}{2} ( \rho + {\cal P}) \theta=0,
\label{ODmom}
\end{eqnarray}
and by the other two equations stemming from the spatial components (i.e. 
$(i=j)$ and $(i\neq j)$) of Eq. (\ref{E1}):
\begin{eqnarray}
&&   \phi'  + \biggl({\cal H} + 2 \frac{{\cal H}'}{{\cal H}}\biggl)- 
\frac{a^2}{2 {\cal H}} \biggl[\delta p - 3 \frac{{\cal H}}{a} \delta \xi - \frac{\overline{\xi}}{a}(\theta - 3 {\cal H}\phi)\biggr] =0 .
\label{OD3}\\
&& B' + 2 {\cal H} B + \phi =0.
\label{OD4}
\end{eqnarray}
In Eq. (\ref{ODham}) the global energy and pressure density 
fluctuations (i.e. $\delta\rho$ and $\delta p$) have been introduced. 
As is clear from Eqs. (\ref{ODham})--(\ref{OD4}),
one of the advantages of the off-diagonal formulation is the absence 
of second time derivatives of the metric fluctuations. Strictly 
speaking the evolution equations of $\theta_{\rm a}$ and $\theta_{\rm b}$ 
should be added to the system. However, they are only relevant 
for typical length scales smaller than the Hubble radius at a given time. 
Since we are interested in the opposite regime, these equations will be 
omitted, but they will be discussed elsewhere in their full generality (see \cite{maxlon}). 

By combining Eqs. (\ref{ODham}) and (\ref{OD3}) the evolution equation for $\zeta$ can be easily obtained; it is given by 
\begin{eqnarray}
 \dot{\zeta} = - \frac{3}{2} \frac{H}{\dot{H}} [ (\dot{\overline{\xi}}_{\rm a} + \dot{\overline{\xi}}_{\rm b})\zeta + 
 H (\delta \xi_{\rm a} + \delta \xi_{\rm b})] + \frac{\dot{\rho}_{\rm a}}{ 2 \dot{H}} ( c_{\rm s\, b}^2 - c_{\rm s\, a}^2)  ( \zeta_{\rm a} - \zeta),
\label{evolzeta}
\end{eqnarray}
where we passed, for later convenience, from the conformal time 
coordinate $\tau$ to the cosmic time coordinate $t$ (i.e. $d t = a(\tau) d\tau$).

Equations (\ref{dca}) and (\ref{dcb}) lead to the evolution 
equations of $\zeta_{\rm a}$ and $\zeta_{\rm b}$ whose explicit form is 
given by
\begin{eqnarray}
\dot{\zeta}_{\rm a} +
\biggl[ \frac{\dot{q}_{\rm a}}{q_{\rm a}}
 + ( 3 H + \overline{\Gamma}) ( 1 + c_{\rm s\, a}^2) 
 \biggr] \zeta_{\rm a} 
 + \frac{9}{q_{\rm a}} [ H^2 \delta \xi_{\rm a} - \overline{\xi}_{\rm a} 
\dot{H} \zeta ] = \frac{ p_{\rm a} + \rho_{\rm a}}{q_{\rm a}} \overline{\Gamma} \biggl[
\delta_{\Gamma} + \frac{\dot{H}}{H^2} \zeta \biggr],
\label{evolzetaa}
\end{eqnarray}
and by 
\begin{eqnarray}
&& \dot{\zeta}_{\rm b}  + \biggl[ \frac{\dot{q}_{\rm b}}{q_{\rm b}} + 3 H ( 1 + c_{\rm s\, b}^2) \biggr] \zeta_{\rm b} - \overline{\Gamma} \frac{\dot{q}_{\rm a}}{q_{\rm b}} 
( 1 + c_{\rm s\, a}^2) \zeta_{\rm a}  + \frac{9}{q_{\rm b}} [ H^2 \delta \xi_{\rm b} - \dot{H} \overline{\xi}_{\rm b} \zeta ] 
\nonumber\\
&& = - \frac{p_{\rm a} + \rho_{\rm a}}{q_{\rm b}} \overline{\Gamma} \biggl[ \delta_{\Gamma} + \frac{\dot{H}}{H^2} \zeta \biggr].
\label{evolzetab}
\end{eqnarray}
where 
\begin{equation}
q_{\rm a} = \frac{\dot{\rho}_{\rm a}}{H},\,\,\,\,\,\,\,\,\,\,\,
q_{\rm b} = \frac{\dot{\rho}_{\rm b}}{H}.
\end{equation}
Various identities can be used 
to bring  Eqs. (\ref{evolzeta}), (\ref{evolzetaa}) and 
(\ref{evolzetab}) to slightly different (but equivalent) forms. In particular:
\begin{itemize}
\item{} using Eq. (\ref{curvdef}), we can always trade the combinations
 $(\zeta_{\rm a} - \zeta)$ and $(\zeta_{\rm b} - \zeta)$ for $\dot{\rho}_{\rm a}/\dot{\rho}
 (\zeta_{\rm a} - \zeta_{\rm b})$ and $\dot{\rho}_{\rm b}/\dot{\rho}
 (\zeta_{\rm a} - \zeta_{\rm b})$;
 \item{}  according to Eq. (\ref{entrdef}), $(\zeta_{\rm a} - \zeta_{\rm b})= - {\cal S}/3$;
 \item{} by virtue of the background equations (\ref{Eb1}) and (\ref{Eb2}), $\dot{H}/H= \dot{\rho}/(2 \rho)$;
 \item{} if the inviscid component of each fluid of the mixture is a perfect 
 fluid, then $c_{\rm s,\,a}^2 = w_{\rm a}$ and $c_{\rm s,\, b}^2 = w_{\rm b}$;
 \item{} finally the background evolution of each fluid, i.e. Eqs. (\ref{bconsa}) and (\ref{bconsb}), 
 may always be employed to obtain equivalent forms of the above equations.
 \end{itemize}
Specific limits of Eqs. (\ref{evolzeta})--(\ref{evolzetab}) will now be 
reproduced.
In the limit $ \overline{\xi}_{\rm a} = \overline{\xi}_{\rm b} = 0$, 
with $\delta_{\Gamma} = \delta\xi_{\rm a} = \delta\xi_{\rm b} =0$ and 
$\dot{\overline{\Gamma}} =0$, 
Eqs. (\ref{evolzeta})--(\ref{evolzetab}) read
\begin{eqnarray}
&& \dot{\zeta} = - \frac{H \, \dot{\rho}_{\rm a} 
\, \dot{\rho}_{\rm b}}{ \dot{\rho}^2} (w_{\rm b} - w_{\rm a}) {\cal S},
\label{wm1}\\
&& \dot{\zeta}_{\rm a} = \frac{\overline{\Gamma}}{6} (w_{\rm a} + 1) 
\frac{\dot{\rho}_{\rm b} \, \rho_{\rm a}}{\dot{\rho}_{\rm a}} {\cal S},
\label{wm2}\\
&& \dot{\zeta}_{\rm b} =0.
\label{wm3}
\end{eqnarray}
In the case $w_{\rm a} =0$ and $w_{\rm b} =1/3$, Eqs.
 (\ref{wm1})--(\ref{wm3}) coincide with the set of equations used in Ref. 
 \cite{MWU} to describe the radiative decay of a massive curvaton whose 
 effective pressure, at the oscillatory stage, reproduces that
 of dusty matter, i.e. $w_{\rm a}=0$. It is then clear, taking the difference 
 between Eqs. (\ref{wm3}) and (\ref{wm2}), that the evolution equation of entropy 
 perturbations 
 \begin{equation}
 \dot{\cal S} = \frac{\overline{\Gamma}}{2} (w_{\rm a} + 1) \frac{\dot{\rho}_{\rm a} \rho_{\rm a}}{\rho \dot{\rho}_{\rm b}} \biggl( 1 - \frac{\dot{\rho}_{\rm b}^2}{\dot{\rho_{\rm a}}^2} - 2 \frac{\rho}{\rho_{\rm a}} \biggr) {\cal S}
 \end{equation}
 is homogeneous and does not contain any $\zeta$-dependent source term.
 
 Sticking to the case of the radiative decay of a dusty fluid, but including 
 the fluctuations of the decay rate, the following system of evolution
equations 
\begin{eqnarray}
&& \dot{\zeta} = \frac{\dot{\rho}}{6 \dot{H}}  ( \zeta_{\rm a} - \zeta), 
\label{gdvz1}\\
&& \dot{\zeta}_{\rm a} - \frac{\dot{g}_{\rm a}}{g_{\rm a}} \zeta_{\rm a}= - g_{\rm a} 
\biggl( \delta_{\Gamma} + \frac{\dot{H}}{H^2} \zeta \biggr),
\label{gdvz2}
\end{eqnarray}
 can be derived   from Eqs. (\ref{evolzeta})--(\ref{evolzetab}) when $w_{\rm b} =1/3$ and $w_{\rm a}=0$. 
 In Eqs. (\ref{gdvz1})  and (\ref{gdvz2}), $g_{\rm a} = -H \rho_{\rm a}/\dot{\rho}_{\rm a}$. Equations (\ref{gdvz1}) and (\ref{gdvz2})
 describe the situation discussed in Ref. \cite{gdvz}, where the dynamics 
 of the inflaton with inhomogeneous decay rate has been discussed (see, for instance, also \cite{kofman,postma,averdi,mazumdar}). If the spatial fluctuations of the decay rate are not a function of the local energy density of the mixture, curvature fluctuations may be generated for length scales larger than the  Hubble radius.
 
Consider now the case where the $\overline{\Gamma}$ is constant,
 the decay is homogeneous (i.e. $\delta_{\Gamma} =0$), but 
 $\xi_{\rm a} = \xi_{\rm a}(\rho_{\rm a})$ and $\xi_{\rm b} = \xi_{\rm b}(\rho_{\rm b})$. This occurrence implies that 
 \begin{equation} 
 \delta \xi_{\rm a} = - \frac{\dot{\overline{\xi}}_{\rm a}}{H} \zeta_{\rm a}, \,\,\,\,\,\,\,\,\,\,\,\, \delta \xi_{\rm b} = - \frac{\dot{\overline{\xi}}_{\rm b}}{H} \zeta_{\rm b}. 
 \label{dxi1}
 \end{equation}
Hence, from Eqs. (\ref{evolzeta})--(\ref{evolzetab}), we obtain, respectively
\begin{eqnarray}
&& \dot{\zeta} = - \frac{\dot{\rho}_{\rm b}}{\dot{\rho}} \biggl[ \frac{H\,\dot{\rho}_{\rm a}}{\dot{\rho}} (w_{\rm b} -w_{\rm a}) + 
\frac{\dot{\rho}}{4 \rho} \biggl( \dot{\overline{\xi}}_{\rm a} - \frac{\dot{\rho}_{\rm a}}{\dot{\rho}_{\rm b}} \dot{\overline{\xi}}_{\rm b}\biggr)\biggr]\,\, {\cal S},
\label{zetax1}\\
&& \dot{\zeta}_{\rm a} = \frac{\dot{\rho}_{\rm b}}{6\,\rho\,\dot{\rho}_{\rm a}} [\overline{\Gamma} (w_{\rm a} + 1)\rho_{\rm a} + 9 H^2 \overline{\xi}_{\rm a}] \,\, {\cal S},
\label{zetax2}\\
&& \dot{\zeta}_{\rm b} = - \frac{\dot{\rho}_{\rm a}}{3 \dot{\rho}_{\rm b}} \biggl[ 
\overline{\Gamma}( w_{\rm a} + 1) \biggl( 1 - \frac{\rho_{\rm a}}{2 \rho}\biggr) +
\frac{3}{2} \overline{\xi}_{\rm b} \biggl] \,\,{\cal S}.
\label{zetax3}
\end{eqnarray}
Again, in this case, it can be easily argued that the evolution of entropy 
fluctuations obeys a homogeneous equation in ${\cal S}$. In fact, combining 
Eqs. (\ref{zetax2}) and (\ref{zetax3}) it is possible to obtain:
\begin{equation}
\dot{{\cal S}} = - \biggl[ \frac{\dot{\rho}_{\rm a} \rho_{\rm a}}{2 \rho \dot{\rho}_{\rm b}} \overline{\Gamma} (w_{\rm a} + 1) \biggl( 1 - \frac{\dot{\rho}_{\rm b}^2}{\dot{\rho}_{\rm a}^2} - 2 \frac{\rho}{\rho_{\rm a}} \biggr) + \frac{3}{2} \biggl( \overline{\xi}_{\rm a} + \frac{\dot{\rho}_{\rm b}^2}{\dot{\rho}_{\rm a}^2} \overline{\xi}_{\rm b} \biggr) \biggr] \,\, {\cal S}.
\label{S1}
\end{equation}
This conclusion can be, however, evaded if $\xi_{\rm a} $ and $\xi_{\rm b}$ 
are functions both of $\rho_{\rm a}$ and $\rho_{\rm b}$, i.e. 
$\xi_{\rm a}= \xi_{\rm a}(\rho_{\rm a},\rho_{\rm b})$ and 
$\xi_{\rm b}= \xi_{\rm b}(\rho_{\rm a},\rho_{\rm b})$. In this case
\begin{equation}
\delta \xi_{\rm a} =  - \frac{\dot{\overline{\xi}}_{\rm a}}{H}( \zeta_{\rm a} + \zeta_{\rm b}),\,\,\,\,\,\,\,\,\,\,\delta \xi_{\rm b} =  - \frac{\dot{\overline{\xi}}_{\rm b}}{H}( \zeta_{\rm a} + \zeta_{\rm b}).
\label{dxi2}
\end{equation}
Thus, in the situation described by Eq. (\ref{dxi2}), Eq. (\ref{S1})
will inherit two extra terms at the right-hand side, i.e. 
\begin{equation}
-\frac{9 \,H^2}{\dot{\rho}_{\rm b}\,\dot{\rho}_{\rm a}} \biggl( \dot{\rho}_{\rm a} \dot{\overline{\xi}}_{\rm b}  \zeta_{\rm a} - \dot{\rho}_{\rm b} \dot{\overline{\xi}}_{\rm a} \zeta_{\rm b} \biggr),
\end{equation}
which cannot be recast, for generic $\xi_{\rm a}$ and $\xi_{\rm b}$, in a single term proportional to ${\cal S}$.

\begin{figure}
\begin{center}
\begin{tabular}{|c|c|}
      \hline
      \hbox{\epsfxsize = 7 cm  \epsffile{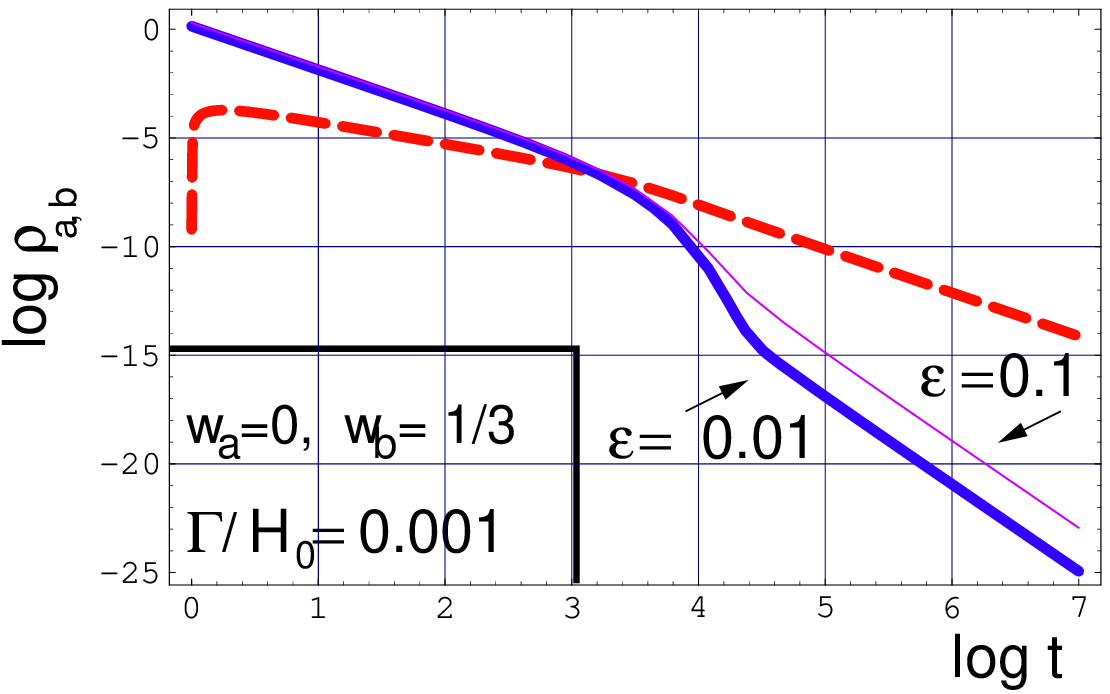}} &
      \hbox{\epsfxsize = 7 cm  \epsffile{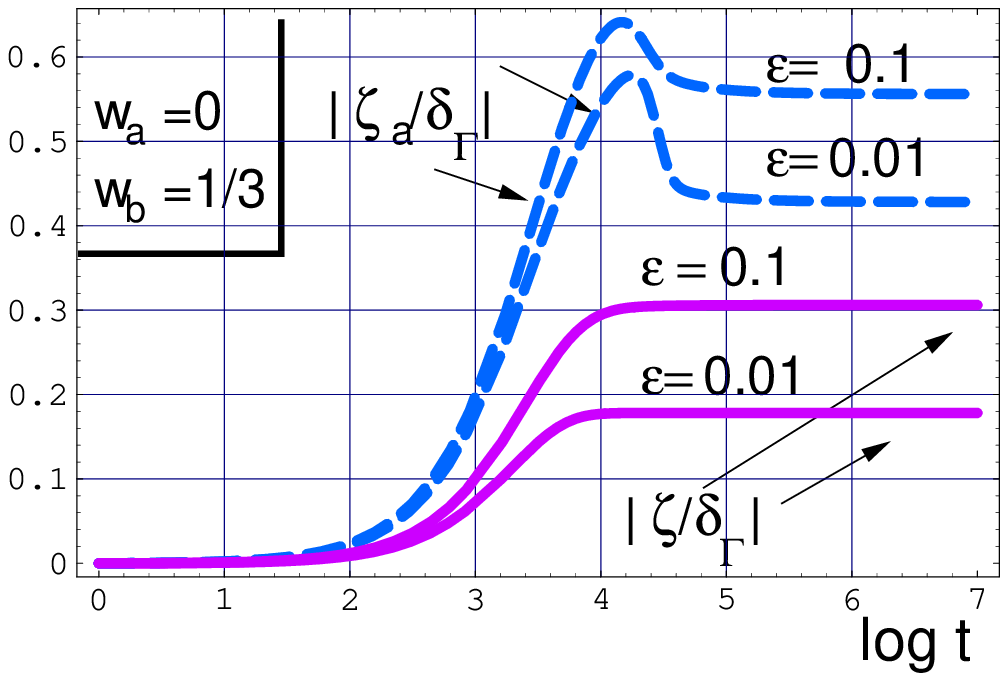}}\\
      \hline
\end{tabular}
\end{center}
\caption[a]{The evolution of the background (left-hand plot) and of the fluctuations (right-hand plot) is illustrated. 
The parameters of the mixture are fixed in such 
a way that $w_{\rm a} =0$, $w_{\rm b} = 1/3$, $\Gamma/H_{0} \sim 10^{-3}$, 
$\xi_{\rm a} = \epsilon \sqrt{\rho_{\rm a}}$, $\xi_{\rm b} =0$; $H_{0}$ denotes the value of the Hubble parameter at the initial integration time. In the left plot, 
the dashed curve represents the evolution of the energy density of the decay products (radiation) while the full  lines represent the 
evolution of the energy density of the decaying component for different 
values of $\epsilon$. In the right plot hand side the dashed curves 
illustrate the behaviour of $|\zeta_{\rm a}|$ while the full lines represent 
the evolution of $|\zeta|$. Both $|\zeta|$ and $|\zeta_{\rm a}|$ are given in units of $\delta_{\Gamma}$.}
\label{F1}
\end{figure}

A relevant issue to be addressed concerns the phenomenological 
viability of  interacting viscous mixtures. Consider, for 
instance, a  model where the decay rate is constant but inhomogeneous (i.e. 
$\delta_{\Gamma} \neq 0$) and $ \xi_{\rm a} = \epsilon \sqrt{\rho_{\rm a}}$ (where $\epsilon$ is constant). The viscosity coefficient 
of the b-fluid vanishes, i.e. $\xi_{\rm b} =0$.   This model describes the situation where the a-fluid is initially dominant and characterized by 
a viscosity proportional to $\epsilon$. Furthermore, if we want  
the Universe to be expanding, we must also require $\epsilon < (w_{\rm a} + 1)/\sqrt{3}$. The a-fluid will start its decay for a typical cosmic time
$t_{\Gamma} \sim \overline{\Gamma}^{-1}$, and then the background 
will be dominated by the b-fluid while the energy density of the a-fluid, i.e. 
$\rho_{\rm a}$, will decay exponentially. Also the background viscosity 
will decay exponentially, since $\overline{\xi}_{\rm a} = \epsilon \sqrt{\rho_{\rm a}}$.  These aspects are illustrated in Fig. \ref{F1} (plot at the left-hand side)
where, for two different values of $\epsilon$, the common logarithm 
(i.e. the logarithm to base 10) of $\rho_{\rm a}$ and $\rho_{\rm b}$ are reported.
From the point of view of the background, this 
model is perfectly viable and it leads to a final stage of expansion 
dominated by the b-fluid. To make the example even more explicit, 
one can think of the situation where the a-fluid is given by dust (i.e. 
$w_{\rm a}=0$) or stiff matter (i.e. $w_{\rm a} = 1$). The b-fluid 
may be taken, for instance, to coincide with radiation (i.e. $w_{\rm b} = 1/3$).

The dynamics of curvature fluctuations may be described, for practical 
reasons, by expressing the evolution equations in terms of $\zeta$ and $\delta_{\rm a}$, i.e. the curvature fluctuations and the density contrast 
of the a-fluid. Given the relations  (\ref{curvdef}) and (\ref{ztod}) 
$\zeta_{\rm b}$, $\zeta_{\rm a}$ and $\delta_{\rm b}$ can 
always be obtained as linear combinations (with background-dependent coefficients) of $\zeta$ and $\delta_{\rm a}$. From Eqs. (\ref{dca}) and 
(\ref{evolzeta})
the relevant evolution equations can be written as 
\begin{eqnarray}
&& \dot{\zeta} = - \frac{1}{ 4 \, \dot{H}} \biggl[ 3 \frac{H\,\epsilon\,}{\sqrt{\rho_{\rm a}}} + 2 (w_{\rm b} - w_{\rm a})\biggr] ( 
\dot{\rho}_{\rm a} \zeta + H\, \rho_{\rm a}\, \delta_{\rm a}),
\label{zexa}\\
&& \dot{\delta}_{\rm a} + \frac{9\,\epsilon \, H^2}{2 \sqrt{\rho_{\rm a}}} \delta_{\rm a} + \frac{\dot{H}}{H^2} \biggl[ \frac{9 \,\epsilon \,H^2}{\sqrt{\rho_{\rm a}}} + \overline{\Gamma} (w_{\rm a} + 1)\biggr] \zeta = - 
\overline{\Gamma} (w_{\rm a} + 1) \delta_{\Gamma}.
\label{zexb}
\end{eqnarray}
Equations (\ref{zexa}) and (\ref{zexb}) describe the evolution of $\zeta$ and $\delta_{\rm a}$ for typical wavelengths larger than the Hubble radius.
Initial conditions of the system are then set by requiring $\zeta(t_{0}) =0$ 
and $\delta_{\rm a}(t_0) = \delta_{\rm b}=0$, where $t_{0}$ is the initial integration time.
From Fig. \ref{F1} (plot at the right-hand side) the evolution is such that 
curvature fluctuations grow from $0$ to a value proportional to $\delta_{\Gamma}$, i.e. proportional to the fluctuations of the decay rate 
over length scales larger than the Hubble radius. The final asymptotic value 
of $\zeta$ can be determined analytically and it turns out to be 
\begin{equation}
|\zeta_{\rm final}| \simeq \frac{1}{6} \biggl(\frac{ 1 + 3 \sqrt{3}\,\epsilon}{ 1 - \sqrt{3}\, \epsilon} \biggr).
\label{as}
\end{equation}
In the limit $\epsilon \to 0$, the results reproduce the findings of Ref. \cite{gdvz}
leading to a Bardeen potential $|\Psi_{\rm final}| \simeq \delta_{\Gamma}/9$, which implies $|\zeta_{\rm final}| \simeq \delta_{\Gamma}/6$ by using the 
well-known relation of $\zeta$ and $\Psi$ in a radiation-dominated phase (see,
for instance, \cite{maxrev}). 
In the example discussed so far the values of $\delta_{\Gamma}$ have been 
taken in the ranges $10^{-6}$--$10^{-9}$.  

The class of examples reported so far can be generalized in various 
ways. Different barotropic indices for the fluids of the mixture 
can be studied. Equation (\ref{as}) can then be generalized to the cases 
of generic $w_{\rm a}$ and $w_{\rm b}$. Furthermore, the functional 
dependence of the viscosity coefficients can be chosen to be different.
Possible generalizations will be present elsewhere \cite{maxlon}.
We would like to point out that the simple examples presented here 
may be made more realistic by thinking that a dust fluid is an effective description of a scalar field oscillating in a quadratic potential \cite{turner}.
Thus, the simple fluid model of a dust fluid decaying into radiation
has been used \cite{gdvz} (with some caveats \cite{mazumdar}) 
to infer some properties of the inflaton decay when 
the inflaton decay rate is not homogeneous. 
If the inhomogeneous decay occurs after an  inflationary phase at 
low curvature (i.e. $H_{\rm inf} \ll 10^{-6} M_{\rm P}$), 
it is plausible to argue that the spectrum 
of $\delta_{\Gamma}$ may be converted into the spectrum of $\zeta$ for 
typical frequencies smaller than the Hubble rate. We are not interested 
here in supporting a specific model of inhomogeneous reheating. The 
purpose of the examples discussed so far is purely illustrative. However, the 
 lesson to be drawn is that bulk viscous stresses may play a relevant role.

In the present paper, various results have been achieved. 
First of all, the concept of interacting viscous mixtures has been 
introduced, i.e.  a mixture of interacting fluids with viscous 
corrections. In this framework, the coupled evolution of curvature and entropy 
fluctuations has been derived in the case where both the decay rate and the 
bulk viscosity coefficients are allowed to fluctuate over typical length scales 
larger than the Hubble radius. Different situations have been 
systematically discussed. If the decay rate is constant and homogeneous, 
with bulk viscosities that depend separately
 on the energy density of each fluid of the mixture, the evolution 
 of entropy fluctuations obeys a source-free evolution equation. 
 If, on the contrary, the bulk viscosity has a more general dependence 
 on the energy densities of the fluids composing the mixture, the 
 evolution equations of the entropy perturbations may inherit a 
 source term that involves, in one way or  another, curvature 
 fluctuations. In similar terms, if the decay rate is allowed to fluctuate 
 without being a function of the local density of the fluid, entropy 
 fluctuations will not obey a source-free equation.


\begin{thebibliography}{99}

\bibitem{KS} H. Kodama and M. Sasaki, Prog. Theor. Phys. Suppl. {\bf 78}, 1 (1984).

\bibitem{bst} J. Bardeen, P. Steinhardt, and M. Turner, Phys. Rev. D {\bf 28}, 679 (1983).

\bibitem{weinberg} S.~Weinberg, Phys.\ Rev.\ D {\bf 67}, 123504 (2003).

\bibitem{maxrev} M. Giovannini, {\it Theoretical tools for CMB physics}, CERN-PH-TH/2004-140, astro-ph/0412601.

\bibitem{turok} M.~Bucher, K.~Moodley and N.~Turok, Phys.\ Rev.\ D {\bf 62}, 083508 (2000).

\bibitem{barrow1}   J.~D.~Barrow, Nucl.\ Phys.\ B {\bf 310}, 743 (1988).

\bibitem{barrow2} J. Barrow, Phys. Lett. B {\bf 180} , 335 (1986).

\bibitem{barrow3} J. Barrow, Phys. Lett. B {\bf 187}, 12 (1987).

\bibitem{maxoff}  R.~Brustein, M.~Gasperini, M.~Giovannini, V.~F.~Mukhanov and G.~Veneziano,
  Phys.\ Rev.\ D {\bf 51}, 6744 (1995).

\bibitem{hoff}  J. Hwang, Astrophys. J.  {\bf 375}, 443 (1990).

\bibitem{hwang}
  J.~c.~Hwang and H.~Noh,  Class.\ Quant.\ Grav.\  {\bf 19}, 527 (2002).

\bibitem{malikwands} K.~A.~Malik and D.~Wands,  JCAP {\bf 0502}, 007 (2005).

\bibitem{maxlon} M. Giovannini, {\it Imperfect cosmological perturbations} (in progress).

\bibitem{bp} R. Brandenberger, R. Kahn, and W. Press, Phys. Rev. D {\bf 28}, 1809 (1983).

\bibitem{lyth} D. H. Lyth, Phys. Rev. D {\bf 31}, 1792 (1985). 

\bibitem{MWU}  K.~A.~Malik, D.~Wands and C.~Ungarelli,
  Phys.\ Rev.\ D {\bf 67}, 063516 (2003).

\bibitem{gdvz} G.~Dvali, A.~Gruzinov and M.~Zaldarriaga,
  Phys.\ Rev.\ D {\bf 69}, 023505 (2004).

\bibitem{kofman} L.~Kofman, arXiv:astro-ph/0303614.

\bibitem{postma}  M.~Postma,  JCAP {\bf 0403}, 006 (2004).

\bibitem{mazumdar} A. Mazumdar and M. Postma, Phys. Lett. B {\bf 573} 5 (2003), [Erratum-ibid. {\bf 585}, 295 (2004)]

\bibitem{averdi} R.~Allahverdi,  Phys.\ Rev.\ D {\bf 70}, 043507 (2004).

\bibitem{turner} M.~S.~Turner,
  Phys.\ Rev.\ D {\bf 28}, 1243 (1983).

\end{thebibliography}
\end{document}